\documentstyle[amssymb,12pt]{article}
\input{tcilatex}

\begin{document}

\baselineskip=30pt

\begin{titlepage}

\vskip1in

\begin{center}
\LARGE{\bf Renormalization Conditions and the Sliding Scale
 in the Implicit Regularization Scheme: A Simple Connection.}
\end{center}

\vskip1.0cm

\begin{center}
\Large{A. Brizola$^{1}$  \,\, S. R. Gobira$^{1}$ \,\, Marcos Sampaio $^{1}$
\,\, M. C. Nemes$^{1,2}$}\\ 
\end{center} 

\vskip1.0cm
\footnotesize{
\begin{center}
$(1)$Federal University of Minas Gerais\\
Physics Department - ICEx\\
P.O. BOX 702, 30.161-970, Belo Horizonte - MG - Brazil\\
\vskip1.0cm
$(2)$ Institute for Physics - S\~ao Paulo University\\
P.O.Box 66 318 - CEP 05315-970\\
S\~ao Paulo - SP - Brazil\\

\end{center}
\vskip0.5cm
\begin{center}
{\it {brizola@fisica.ufmg.br, gobira@fisica.ufmg.br, msampaio@fisica.ufmg.br, carolina@fisica.ufmg.br}}
\end{center}}
\newpage
\begin{abstract}
\noindent
We describe in detail how a sliding scale is introduced in the renormalization of a QFT according to  integer dimensional implicit regularization scheme. We show that since no regulator needs to be specified at intermediate steps of the calculation, the introduction of a mass scale is a direct consequence of a set of renormalization conditions. As an illustration the one loop $\beta$-function for $QED$ and $\lambda \varphi^4$ theories are derived. They are given in terms of derivatives of appropriately sistematized functions (related to definite parts of the amplitudes) with respect to a mass scale $\mu$. Our formal scheme can be easily generalized for higher loop calculations.

\end{abstract}
%%%%%%%%%%%%%%%%%%%%%%%%%%%%%%%%%%%%%%%%%%%%%%%%%%%%%%%%%%%%%
\end{titlepage}

\section{\protect\bigskip Introduction}

In dealing with quantum field theories usually divergent quantities (Green
functions, Feynman integrals, etc.) are found in some large energy region.
These large momenta correspond to short distance singularities resulting
from badly defined quantities such as products of fields at the same point.
As these elementary Green functions are not well defined functions but
rather distributions and since product of distributions is ill defined, this
leads to the divergences in large momenta aforementioned. These quantities
should, in principle, have physical content. In order to proceed, a
regulator must be imposed in the divergent expressions and, in the process
of renormalization, finite parameters are defined. The renormalization
program has to be a systematic and unambiguously fixed algorithm that
satisfies the fundamental properties of locality and causality \cite
{citbonn1}: it should correspond to the addition of local counterterms to
the Lagrangian density. The predictivity of the theory, that is the ability
to obtain results valid to all orders of perturbation theory such as the
renormalization group equation, rely on these logical conditions. In general
lines, any renormalization procedure involves two steps \cite{bonn}:

{\em 1. A regularization followed by a subtraction procedure;}

{\em 2. A set of renormalization conditions in order to define the
parameters of the theory;}

Step 1 refers to a systematic, uniquely fixed and consistent procedure
powerful enough to investigate to all orders of perturbation theory its
renormalizability, fields, finite parameters and symmetries. A common
feature of regularization schemes is the introduction of \ (at least) one
regularizing parameter (sharp cutoff \ $\Lambda ^{2}$, Pauli-Villars masses $%
m_{i}^{2}$, dimension of space-time $D=4-\epsilon $, ...) and in the process
of subtracting the divergences, the resulting expressions will remain finite
when the regulator is suppressed. Stated generally, the introduction of
regulators is followed by the introduction of mass parameters. The second
step refers to the task of defining the parameters of the theory (fields,
masses, couplings) in a suitable energy scale $\mu $\ in each order of
perturbation \cite{pesk}. This is accomplished by subtracting each primitive
divergence from a specific parameter. In other words, after introduction of
a regularization, the coefficients of the counterterms are completely
determined by renormalization conditions, imposed order by order on the
primitively divergent Green functions. The infinite subtraction is performed
in an energy scale $\mu $ and the study of the behavior of renomalized Green
functions with $\mu $ is an important branch in particle physics, issuing
the renormalization group techniques \cite{4} \cite{5} \cite{6} \cite{7} 
\cite{8} \cite{artigo1}.

In implementing step 1 above the most successful and popular regularization
procedure is the Dimensional Regularization (DR) \cite{tv}. The great
success of DR is mainly due to the fact that it automatically respects gauge
invariance. It is known, however, that it presents problems in
dimensional-dependent theories like chiral or supersymmetric theories. The
proposed alternative, dimensional reduction \cite{dr} is usually employed in
that cases, although inconsistencies may arise at high orders \cite{ho}. In
this context it is most desirable to develop other regularization schemes,
specific to 4-dimensions which preserves the consistency of DR. Recently two
such schemes were proposed, the Differential Regularization and the Implicit
Regularization. The first one is established in coordinate space and the
latter in momentum space. A mass scale is automatically introduced in
Differential Regularization, for dimensional reasons, when the regulated
propagators are defined. The relation between this mass parameter and the
choice of the renormalization point, as in DR, is not direct. As we will
show, since in the Implicit Regularization scheme no specific form of a
regulator need to be specified, the calculation is not contaminated by
regularization parameters in any step. This technique is therefore most
adequate to establish, in a regularization independent way, the relation
between the mass scale and the choice of the renormalization point. This is
the main purpose of the present contribution.

In section II we introduce the Implicit Regularization Technique, the
sliding scale in the renormalization procedure and systematize the finite
contributions of two and three point functions. In section III we discuss
the Renormalization Group within our scheme and derive the $\beta $-function
for QED and $\lambda \varphi ^{4}$. Relation between our approach and the
other current schemes can be found in section IV. \ In section V we compare
differential, dimensional and implicit renormalizations. Final remarks are
in section VI.

\section{The Implicit Regularization Technique and Finite Content of
One-Loop Amplitudes}

In this section we define the Implicit Regularization Technique (IRT) for a
general n-loop calculation. We closely follow ref.\cite{ri3}. The first step
in implementing the IRT is to assume an implicit regularization\footnote{%
The only required condition about the implicit regularization is that it
must be even in the loop momenta and with a connection limit that returns
the original integrand.} whenever a divergence occurs in a Feynman integral.
After taking the Dirac trace (if required), one identifies the divergence
degree of the integrals and manipulates the integrand by means of algebraic
identities until the external momenta dependent parts are isolated solely in
terms of finite contributions. To separate the divergences the following
identity will be used recursively until the last term acquires a negative
degree of divergence in an integration over $k$ in 4 space-time dimensions: 
\begin{eqnarray}
\frac{1}{[(k+p)^{2}-m^{2}]} &=&\sum_{j=0}^{N}\frac{(-1)^{j}(p^{2}+2p\cdot
k)^{j}}{(k^{2}-m^{2})^{j+1}}+  \nonumber \\
&&\frac{(-1)^{N+1}(p^{2}+2p\cdot k)^{N+1}}{(k^{2}-m^{2})^{N+1}}\frac{1}{%
[(k+p)^{2}-m^{2}]}.  \label{ident}
\end{eqnarray}
By convenience we divide the diagrams which contribute to a given order in
two classes: the first which do not contain diagrams which possess two point
functions as subdivergences and in the second class those which do.

Let us start with the first class of diagrams. To show how the procedure
works it is enough to consider a general Feynman amplitude with one external
momentum $p$, one coupling constant $\lambda $ and one mass parameter $m$ .
We work in the $4$-dimensional space-time although the generalization to any
integer dimension is straightforward. We denote by $q$ a sum of internal
momenta $k_{i}$ . The amplitude in question can always be written as

\begin{equation}
\Gamma =\prod_{i=1}^{n}\int_{\Lambda }\frac{d^{4}k_{i}}{(2\pi )^{4}}%
R(p,q,m,\lambda )\left[ \prod_{j=1}^{l}f_{j}(p,q_{j},m^{2})\right]  \label{7}
\end{equation}
where $\Gamma$ represents $1-PI$ diagrams, 
\begin{equation}
f_{j}(p,q_{j},m^{2})=\frac{1}{[(p-q_{j})^{2}-m^{2}]}  \label{8}
\end{equation}
and 
\[
\mbox{l}=\mbox{number of}\,\,\ f\,\,\,\mbox{structures} 
\]
\[
\mbox{n}=\mbox{number of loops.}\cdot 
\]

Note that we have explicitly separated the terms involving the external
momentum in the denominator, from which nonlocal divergent contributions can
arise after integration over the internal momenta. The structure $%
R(p,q,m,\lambda )$ contains all other ingredients of the amplitude such as
coupling constants, results of Dirac traces, and so on.

For simplicity we adopt the following notation 
\begin{equation}
\Gamma =(\Pi R)(\Pi f)  \label{9}
\end{equation}
where 
\begin{equation}
(\Pi R)=\prod_{i=1}^{n}\int_{\Lambda }\frac{d^{4}k_{i}}{(2\pi )^{4}}%
R(p,q,m,\lambda )  \label{10}
\end{equation}
and 
\begin{equation}
(\Pi f)=\prod_{j=1}^{l}f_{j}(p,q_{j},m^{2})\cdot  \label{11}
\end{equation}
As discussed before the source of all possible troubles in the
renormalization process will arise from the structure $(\Pi f).$ Our method
focus attention on these structures. In order to clearly separate finite,
``trivial'' divergences (whose dependence on the external momenta is only a
polynomial) from the nonlocal divergences we use a strategy which is
completely based on the identity (\ref{ident}).

Define the operator $T^{D}$ which acts on {\it each} structure $f$ in the
following way 
\begin{equation}
T^{0}f=\frac{1}{q_{j}^{2}-m^{2}}+\frac{2p.q_{j}-p^{2}}{(q_{j}^{2}-m^{2})}%
\left\{ \frac{1}{[(p-q_{j})^{2}-m^{2}]}\right\}  \label{12}
\end{equation}
\begin{equation}
T^{1}f=\frac{1}{q_{j}^{2}-m^{2}}+\frac{(2p.q_{j}-p^{2})}{%
(q_{j}^{2}-m^{2})^{2}}+\frac{(2p.q_{j}-p^{2})^{2}}{(q_{j}^{2}-m^{2})^{2}}%
\left\{ \frac{1}{[(p-q_{j})^{2}-m^{2}]}\right\}  \label{13}
\end{equation}
\begin{eqnarray}
T^{2}f &=&\frac{1}{q_{j}^{2}-m^{2}}+\frac{(2p.q_{j}-p^{2})}{%
(q_{j}^{2}-m^{2})^{2}}+\frac{(2p.q_{j}-p^{2})^{2}}{(q_{j}^{2}-m^{2})^{3}} 
\nonumber \\
&&+\frac{(2p.q_{j}-p^{2})^{3}}{(q_{j}^{2}-m^{2})^{3}}\left\{ \frac{1}{%
[(p-q_{j})^{2}-m^{2}]}\right\} \cdot  \label{14}
\end{eqnarray}
Note that the action of the operator $T^{D}$ is equivalent to a Taylor
expansion around zero external momentum where the first terms are kept and
the rest of the series is resumed, yielding thus a convenient identity. Note
also that the degree of divergence of the various terms is decreasing.

The procedure we have in mind consists of applying the operation, in a
particular amplitude with the superficial degree of divergence $D$ , to {\it %
each} function $f_{j}$ 
\begin{equation}
T^{D}\Gamma =(\Pi R)\prod_{j=1}^{l}T_{j}^{D}f_{j}(p,q_{j},m^{2})\cdot
\label{15}
\end{equation}
The result of the operation will always have the form 
\begin{equation}
T^{D}f(p,q,m^{2})=f^{div}(p,q,m^{2})+f^{fin}(p,q,m^{2})\cdot  \label{16}
\end{equation}
We define 
\begin{equation}
f^{div}(p,q,m^{2})=\sum_{i=0}^{D}f^{i}(p,q,m^{2})\cdot  \label{17}
\end{equation}
Let us exemplify. Take a quadratically divergent amplitude. To each
contribution of the form 
\[
\frac{1}{(p-q_{j})^{2}-m^{2}} 
\]
we associate

\begin{equation}
f^{0}(q,m^{2})=\frac{1}{q^{2}-m^{2}}  \label{18}
\end{equation}

\begin{equation}
f^{1}(p,q,m^{2})=\frac{2p.q-p^{2}}{(q^{2}-m^{2})^{2}}  \label{19}
\end{equation}
\begin{equation}
f^{2}(p,q,m^{2})=\frac{(2p.q)^{2}}{(q^{2}-m^{2})^{3}}  \label{20}
\end{equation}
and 
\begin{equation}
f^{fin}(p,q,m^{2})=\frac{p^{4}-4p^{2}(p.q)}{(q^{2}-m^{2})^{3}}+\frac{%
(2p.q-p^{2})^{3}}{(q^{2}-m^{2})^{3}[(p-q_{j})^{2}-m^{2}]}\cdot  \label{21}
\end{equation}
The definitions (\ref{18}), (\ref{19}), (\ref{20}), (\ref{21}) are not
unique. It is simply convenient for our purposes. Using these we rewrite the
amplitude as a sum of various contributions. According to our notation 
\begin{equation}
T^{D}\Gamma =(\Pi
R)\prod_{j=1}^{l}[f_{j}^{div}(p,q,m^{2})+f_{j}^{fin}(p,q,m^{2})]\cdot
\label{22}
\end{equation}
In this way we can identify three distinct contributions for the amplitude 
\begin{equation}
T^{D}\Gamma =\Gamma _{fin}^{1}+\Gamma _{local}+\Gamma _{nonlocal}  \label{23}
\end{equation}
where 
\begin{equation}
\Gamma _{fin}^{1}=(\Pi R)\prod_{j=1}^{l}f_{j}^{fin}(p,q,m^{2})\cdot
\label{24}
\end{equation}
The second contribution contains only local divergences and, for some
particular $(\Pi R)$ structure, it can contain finite contributions too. It
is identified as 
\begin{eqnarray}
\Gamma _{local} &=&(\Pi R)\prod_{j=1}^{l}f_{j}^{div}(p,q,m^{2})  \nonumber \\
&=&\Gamma _{fin}^{2}+\Gamma _{local}^{div}\cdot  \label{25}
\end{eqnarray}
These local divergences correspond to counterterms which are characteristic
of the order we are renormalizing. For example, they can have the form 
\begin{equation}
\int_{\Lambda }\frac{d^{4}k}{(2\pi )^{4}}\frac{1}{k^{2}-m^{2}}%
+p^{2}I_{log}(m^{2})+\,\,{\mbox{finite\,\,\, part}}\,\cdot  \label{26}
\end{equation}
The last term in equation (\ref{23}), namely the cross-terms, contain finite
contributions as well as ``nonlocal'' divergences. 
\begin{equation}
\Gamma _{nonlocal}=\Gamma _{fin}^{3}+\Gamma _{nonlocal}^{div}\cdot
\label{27}
\end{equation}
These nonlocal divergence contributions will always appear due to the
divergent subdiagrams (beyond two point functions) contained in the graph.
As we will show next in a particular example, the renormalization of
previous orders will always allow one to cancel these contributions if the
theory is renormalizable. In the present scheme the result is automatic and
follows from the operation we have just defined, in an algebraic manner.
There is no need for graphic representations of relevant contributions,
although it is possible.

The renormalized amplitude say, in $n^{th}$ loop order, can therefore be
defined as 
\begin{eqnarray}
\Gamma _{R}^{(n)} &=&T^{D}\Gamma ^{(n)}-\Gamma _{local}^{div(n)}-\Gamma
_{nonlocal}^{div(n)} \\
&=&\Gamma _{fin}^{1(n)}+\Gamma _{fin}^{2(n)}+\Gamma _{fin}^{3(n)}  \nonumber
\end{eqnarray}
where the contributions $\Gamma _{local}^{div(n)}$ and $\Gamma
_{nonlocal}^{div(n)}$ contain the counterterms typical of\ order n as well
as the counterterms coming from divergent subdiagrams of previous order.
Notice from the equation above that our framework automatically delivers the
counterterms 
\begin{equation}
\Gamma _{CT}^{1}=-\Gamma _{local}^{div}-\Gamma _{nonlocal}^{div}
\end{equation}
and just as in BPHZ , by subtracting off the necessary counterterms leaves
with the finite part of the amplitude, the main difference being that here
the counterterms can be read out of the procedure.

Now we proceed to the second class of diagrams, namely those which contain
two point functions as subdiagrams. Let us call $U$ all the two point
diagrams contained in a given amplitude $\Gamma $. It is easy to see that
they can be factored out inside of the total amplitude in the following
sense 
\begin{equation}
\Gamma =\prod_{all\ \ \Sigma _{j}\ \in \ U}{\cal R}_{j}\Sigma
_{j}^{(l)}(q_{j}^{2})
\end{equation}
where ${\cal R}_{j}$ stands for the remaining pieces in the amplitude, $j$
characterizes a specific two point function, is one of the integration
momenta (but external to $\Sigma _{j}$ ). Now since the operation $%
T^{D}\Gamma $ is an identity, i.e.$T^{D}\Gamma =\Gamma $ we can define the
partially renormalized amplitude (with all two point function subdiagrams
properly renormalized ) as follows 
\begin{equation}
\bar{\Gamma}=\Gamma +\Gamma _{CT}^{2}
\end{equation}
therefore we have 
\begin{equation}
\Gamma _{CT}^{2}=\prod_{all\ \ \Sigma _{j}\ \in \ U}{\cal R}_{j}[\delta
_{j}^{(l)}m^{2}-A_{j}^{(l)}q_{j}^{2}]
\end{equation}
and $\Gamma _{CT}^{2}$ are all counterterms characteristic subdiagrams
involving two point functions. $\delta _{j}^{(l)}m^{2}$ stands for the mass
renormalization and $A_{j}^{(l)}$ for the wave function renormalization. In
order to get the renormalized amplitude of order n from $\bar{\Gamma}$ one
proceeds in the same way as for diagrams of class one defined above. We thus
have 
\begin{eqnarray}
\Gamma _{R} &=&T^{D}\bar{\Gamma}-\bar{\Gamma}_{local}^{div}-\bar{\Gamma}%
_{nonlocal}^{div}  \nonumber \\
&=&\bar{\Gamma}_{fin}^{1}+\bar{\Gamma}_{fin}^{2}+\bar{\Gamma}_{fin}^{3}\cdot
\end{eqnarray}

Summarizing, the amplitudes will be written as the sum of basically
divergent parts (defined in each order of perturbation), terms containing
differences between divergent integrals of the same degree of divergence
(which we will call consistency relations) and finite parts. A word about
the consistency relations\ is in order. An important ingredient of the IRT
are the so called consistency relations expressed by differences between
divergent integrals of the same degree of divergence. It was shown \cite{ri1}
that such consistency relations should vanish in order to avoid ambiguities
related to the various possible choices for the momentum routing in certain
amplitudes involving loops, consistently with gauge invariance. This is an
important feature of dimensional regularization and it can be easily checked
that the consistency relations are readily fulfilled in the framework of
dimensional regularization. Alternatively and more generically we can assign
an arbitrary value to such consistency relations and let general symmetry
properties of the theory or physical constraints determine their value \cite
{ri2}.

Let us now consider the massive $\lambda \varphi _{4}^{4}$ theory \cite{31} 
\begin{equation}
{\cal {L}}_{B}=\frac{1}{2}\partial _{\nu }\varphi _{B}\partial ^{\nu
}\varphi _{B}-\frac{m_{B}^{2}}{2}\varphi _{B}^{2}-\frac{\lambda _{B}}{4!}%
\varphi _{B}^{4}.  \label{lag}
\end{equation}
The index $B$ means bare parameters. In order to renormalize the theory the
multiplicative renormalization constants $z_{\varphi },$ $z_{\lambda },z_{m}$
are introduced 
\begin{equation}
\varphi _{B}=z_{\varphi }^{1/2}\varphi ,
\end{equation}
\begin{equation}
\lambda _{B}=z_{\varphi }^{-2}z_{\lambda }\lambda ,
\end{equation}
\begin{equation}
m_{B}^{2}=z_{\varphi }^{-1}z_{m}m^{2},
\end{equation}
\ Perturbative calculations yield an expansion of $n$-point Green's
function\ $\Gamma ^{(n)}$\ in a {\em conventional defined coupling} 
\begin{equation}
\Gamma ^{(n)}(p^{2})=\sum_{i=0}^{\infty }c_{i}^{(n)}\lambda ^{i},
\label{ser}
\end{equation}
where $\lambda $ is finite and defined in a conventional renormalization
point. Let us define the conventional coupling via the renormalization
conditions: 
\begin{equation}
\Gamma ^{(2)}(p^{2})=m^{2}\,\,\mbox{at}\,\,p^{2}=0  \label{crp1}
\end{equation}
\begin{eqnarray}
\Gamma ^{(4)}(p_{1},p_{2},p_{3},p_{4}) &=&-\lambda  \label{crp2} \\
\mbox\,\,{\ at\ }\,\,p_{i}^{2} &=&0\,\,\mbox{ and }\,\,p_{i}.p_{j}=0
\end{eqnarray}

\begin{equation}
\frac{\partial }{\partial p^{2}}\Gamma ^{(2)}(p^2)=1\,\, \mbox { at }\,\,
p^{2}=0.  \label{crp3}
\end{equation}
The choice of this particular value of the external momenta in (\ref{crp1}),
(\ref{crp2}) and (\ref{crp3}) was guided only for convenience since it
renders specially simple expressions. But it is worthwhile saying that the
very same results would be obtained if the renormalization conditions were
defined in another numerical value of the external momenta \cite{3}. The
most general case is the definition of renormalized parameters on a sliding
scale $\mu $. To accomplish this\ renormalization conditions in a point $\mu 
$ the following conditions must be imposed \cite{a} 
\begin{equation}
\Gamma ^{(2)}(p^2)=-m_{\mu }^{2}\,\, \mbox{ at }\,\, p^{2}=-\mu ^{2}
\label{2p}
\end{equation}
\begin{eqnarray}
\Gamma ^{(4)}(p_{1},p_{2},p_{3},p_{4}) &=&-\lambda _{\mu }  \label{4p} \\
\mbox{ at \ }\,\, p_{i}^{2} &=&\mu ^{2}\,\, \mbox{ and }\,\, p_{i}.p_{j}=-%
\frac{\mu ^{2}}{3},i\neq j.
\end{eqnarray}

\begin{equation}
\frac{\partial }{\partial p^{2}}\Gamma ^{(2)}(p^2)=1\,\, \mbox{ at }\,\,
p^{2}=\mu ^{2}.  \label{df}
\end{equation}
The renormalized coupling (\ref{2p}), the renormalized mass (\ref{4p}) and
the field normalization (\ref{df}) are defined at a sliding scale $\mu $.

As an example, consider the one loop 4-point Green calculation 
\begin{equation}
\Gamma _{1}^{(4)}=-z_{\lambda }\lambda +(z_{\lambda }\lambda )^{2}\left[
\int_{\Lambda }\frac{d^{4}k}{(2\pi )^{4}}\frac{1}{%
[(k+p)^{2}-m^{2}](k^{2}-m^{2})}\right] +O(\lambda ^{3}).  \label{ga4}
\end{equation}
We identify the logarithmically divergent integral 
\begin{equation}
I=\int_{\Lambda }\frac{d^{4}k}{(2\pi )^{4}}\frac{1}{%
[(k+p)^{2}-m^{2}](k^{2}-m^{2})}.  \label{int}
\end{equation}
The symbol $\Lambda $ presupposes an implicit regularization. To separate
the logarithmic divergence according to the IRT, one should apply the $T^{0}$
operator on (\ref{int}) 
\begin{equation}
I=\int_{\Lambda }\frac{d^{4}k}{(2\pi )^{4}}\frac{1}{(k^{2}-m^{2})^{2}}-\int 
\frac{d^{4}k}{(2\pi )^{4}}\frac{p^{2}+2p\cdot k}{%
[(k+p)^{2}-m^{2}](k^{2}-m^{2})^{2}}.  \label{isep}
\end{equation}
The first integral is divergent and the second finite 
\begin{equation}
I=\int_{\Lambda }\frac{d^{4}k}{(2\pi )^{4}}\frac{1}{(k^{2}-m^{2})^{2}}-\frac{%
i}{(4\pi )^{2}}Z_{0}(m^{2},m^{2},p^{2})  \label{int1}
\end{equation}
where 
\begin{equation}
Z_{0}(m^{2},m^{2},p^{2})=\int_{0}^{1}dz\ln \left( \frac{p^{2}z(1-z)-m^{2}}{%
-m^{2}}\right) .  \label{z0}
\end{equation}
In calculating the finite part of (\ref{int}) standard methods have been
used \cite{pork}. Defining $z_{\lambda }$\ in order to cancel the divergence
and imposing the renormalization conditions (\ref{4p}) on (\ref{ga4}) yields
the expansion in the conventional coupling 
\begin{equation}
\lambda _{\mu }=-\lambda -\frac{3}{2}\frac{1}{(4\pi )^{2}}Z_{0}(\mu
^{2},m^{2},m^{2})\lambda ^{2}+O(\lambda ^{3}).  \label{lamb}
\end{equation}
Notice that since no explicit form of a regulator has been used, one can
make immediate contact with other regularizations. The remarkable aspect of (%
\ref{lamb}) is that the dependence on the sliding scale $\mu $ of the
coupling $\lambda _{\mu }$ is entirely concentrated on the $Z_{0}$ function.
In other words, the parameter $\lambda $ is ``fixed'' regarding the sliding
scale $\mu $. This fact points towards a generalization viz., that the very
physical content of a theory is concentrated in finite parts which stems
from an infinite renormalization procedure. Details of calculations of one
loop quantum electrodynamic amplitudes and their associated Ward identities
by using IRT can be found in \cite{ri1}, \cite{ri2} and \cite{ri3}. In what
follow, we present the functions which systematize the finite parts of two
and three point amplitudes and some useful relations between them in some
specific examples.

\subsection{The $Z_{\protect\alpha }$ functions}

The application of the $T$ operator in $n^{th}$-order Green's function
yields finite parts as stated in section 2. In one loop calculations, the
2-point amplitudes with at least two propagators and one external momenta
will be systematized by the dimensionless $Z_{\alpha }$\ functions\footnote{%
The external momenta will restrict to the Euclidean region $p^{2}<0.$} \cite
{ori}: 
\begin{equation}
Z_{\alpha }(p^{2},m_{1}^{2},m_{2}^{2})=\int_{0}^{1}dzz^{\alpha }\ln (\frac{%
p^{2}z(1-z)-(m_{1}^{2}-m_{2}^{2})z-m_{1}^{2}}{-m_{2}^{2}})  \label{zk}
\end{equation}
where $m_{i}^{2}$ stands for a mass parameters, $p^{2}$\ the external
momentum and $\alpha \geq 0$. Usually, 2-point Green functions are restrict
to single mass particles. Taking $m_{1}^{2}=m_{2}^{2}\equiv m^{2}$ the $%
Z_{\alpha }$\ functions assume their most simple form 
\begin{equation}
Z_{\alpha }(p^{2},m^{2},m^{2})=\int_{0}^{1}dzz^{\alpha }\ln (\frac{%
p^{2}z(1-z)-m^{2}}{-m^{2}}).  \label{zkm}
\end{equation}
Eq. (\ref{zkm}) is not restricted to one mass parameter only, since the
follow identity holds 
\begin{equation}
Z_{\alpha }(p^{2},m^{2},M^{2})=Z_{\alpha }(p^{2},m^{2},m^{2})+\frac{1}{%
\alpha +1}\ln (\frac{m^{2}}{M^{2}}),  \label{re}
\end{equation}
where $M^{2}$ stands for another mass parameter. An important aspect of
quantum field theory calculations is the study of Green functions in the
asymptotic region \cite{pesk}. In the limit $p^{2}>>m^{2}$ (\ref{zkm})
becomes 
\begin{equation}
\lim_{p^{2}>>m^{2}}Z_{\alpha }(p^{2},m^{2},m^{2})\rightarrow \frac{1}{%
1+\alpha }\ln (\frac{p^{2}}{m^{2}})  \label{lim}
\end{equation}

Some examples of the use of the $Z_{\alpha }$\ functions are in order.
Consider the quantum electrodynamics bare Lagrangian density \cite{31} 
\begin{equation}
{\cal {L}}=i\bar{\Psi}_{B}\gamma ^{\nu }\partial _{\nu }\Psi _{B}-m_{B}\bar{%
\Psi}_{B}\Psi _{B}-\frac{1}{4}F_{B}^{\alpha \beta }F_{\alpha \beta
}^{B}+e_{B}\bar{\Psi}_{B}\gamma ^{\nu }A_{\nu }^{B}\Psi _{B},
\end{equation}
where 
\begin{equation}
F_{B}^{\alpha \beta }\equiv \partial ^{\alpha }A_{B}^{\beta }-\partial
^{\beta }A_{B}^{\alpha }
\end{equation}
and 
\begin{equation}
\bar{\Psi}_{B}\equiv \Psi _{B}^{\dagger }\gamma ^{0}.
\end{equation}
Multiplicative renormalization constants yield renormalized parameters 
\begin{equation}
A_{\nu }^{B}=\sqrt{z_{3}}A_{\nu },
\end{equation}
\begin{equation}
\Psi _{B}=\sqrt{z_{2}}\Psi ,
\end{equation}
\begin{equation}
e_{B}=\frac{z_{1}}{z_{2}\sqrt{z_{3}}}e
\end{equation}
and 
\begin{equation}
m_{B}=\frac{z_{0}}{z_{2}}m.
\end{equation}
Canonical renormalization conditions define renormalized on-shell parameters 
\footnote{%
Here we use the notation $\not{p}\equiv \gamma _{\mu }p^{\mu }.$}, viz. 
\begin{equation}
\Sigma (\not{p}=m)=0,  \label{zig}
\end{equation}
\begin{equation}
\frac{d}{d\not{p}}\Sigma (\not{p})\mid _{\not{p}=m}=0  \label{dzig}
\end{equation}
\begin{equation}
\Pi (q^{2}=0)=0,  \label{vac}
\end{equation}
and 
\begin{equation}
-ie\Gamma ^{\nu }(p-q=0)=-ie\gamma ^{\nu },  \label{gamm}
\end{equation}
where (\ref{zig}) fixes the electron mass $m$, (\ref{dzig}) and (\ref{vac})
fix the residues of the electron and photon propagators at 1 respectively
and (\ref{gamm}) fixes the electron charge to be $e$. Though quantum
electrodynamics has a ``natural'' definition of the parameters $e$ and $m$,
renormalization conditions can be imposed in order to define the parameters
on a sliding scale $\mu .$ For instance, define (\ref{gamm}) off-shell: 
\begin{equation}
-ie\Gamma ^{\nu }(p-q=0)=-ie\gamma ^{\nu }  \label{fos}
\end{equation}
where $\mu $ stands for a sliding scale. Multiplicative renormalization
yields a renormalized Lagrangian density whose parameters were defined in a
conventional renormalization point 
\begin{equation}
{\cal {L}}=-\frac{1}{4}F_{\mu \nu }F^{\mu \nu }-\bar{\Psi}[\gamma ^{\mu
}(\partial _{\mu }+ieA_{\mu })+m]\Psi  \label{qed}
\end{equation}
Perturbative calculations on (\ref{qed})\ yield 1-loop first order
self-energy and vacuum polarization tensor. The first is given by 
\begin{equation}
-i\Sigma (p)=-e^{2}\int_{\Lambda }\frac{d^{4}k}{(2\pi )^{4}}\frac{\gamma
_{\mu }(\gamma ^{\mu }p_{\mu }-\gamma ^{\mu }k_{\mu }+m)\gamma ^{\mu }}{%
[(p-k)^{2}-m^{2}](k^{2}-m^{2})}.  \label{aae}
\end{equation}
The use of IRT yields \cite{ri0} 
\begin{eqnarray}
\Sigma (p) &=&-ie^{2}(\gamma ^{\mu }p_{\mu }-4m)I_{\ell }(m^{2})+ \\
&&+\frac{e^{2}}{8\pi ^{2}}[(\gamma ^{\mu }p_{\mu }-2m)Z_{0}(\kappa
^{2},m^{2},p^{2})+\gamma ^{\mu }p_{\mu }Z_{1}(\kappa ^{2},m^{2},p^{2})],
\label{aee}
\end{eqnarray}
where $\kappa $ is an infrared cut-off, $m$ the electron mass and $p$ the
external momentum. In (\ref{aee}) we separate the amplitude in a basic
divergent integral with logarithmic divergence (in the limit $\Lambda
\rightarrow \infty $): 
\begin{equation}
I_{\ell }(m^{2})=\int_{\Lambda }\frac{d^{4}k}{(2\pi )^{4}}\frac{1}{%
(k^{2}-m^{2})^{2}},  \label{ilog}
\end{equation}
and a finite part systematized by two $Z_{\alpha }$ functions 
\begin{equation}
Z_{0}(\mu ^{2},m^{2},p^{2})=\int_{0}^{1}dz\ln (\frac{p^{2}z(1-z)+(\mu
^{2}-m^{2})-\mu ^{2}}{-m^{2}})  \label{z0m}
\end{equation}
and 
\begin{equation}
Z_{1}(\mu ^{2},m^{2},p^{2})=\int_{0}^{1}dzz\ln (\frac{p^{2}z(1-z)+(\mu
^{2}-m^{2})-\mu ^{2}}{-m^{2}}).  \label{z1m}
\end{equation}

Another example is the vacuum polarization tensor 
\begin{equation}
-i\Pi _{\mu \nu }(q)=-e^{2}\int_{\Lambda }\frac{d^{4}k}{(2\pi )^{4}}%
Tr\left\{ \frac{\gamma _{\nu }(\gamma ^{\mu }k_{\mu }-\gamma ^{\mu }q_{\mu
}+m)\gamma _{\mu }(\gamma ^{\mu }k_{\mu }+m)}{[(k-q)^{2}-m^{2}](k^{2}-m^{2})}%
\right\} ,  \label{tpv}
\end{equation}
which yields \cite{ri0} according to the IRT 
\begin{eqnarray}
\Pi _{\mu \nu }(q) &=&-\frac{4}{3}e^{2}\frac{1}{(4\pi )^{2}}(q_{\mu }q_{\nu
}-q^{2}g_{\mu \nu })\left[ \frac{1}{q^{2}}%
(q^{2}+2m^{2})Z_{0}(m^{2},m^{2},q^{2})+\frac{1}{3}\right] + \\
&&-\frac{4}{3}e^{2}(q_{\mu }q_{\nu }-q^{2}g_{\mu \nu })I_{\ell }(m^{2}),
\label{tp}
\end{eqnarray}
where finite and divergent contributions are clearly separated.

Let us now consider the functions associated with three point Green's
functions.

\subsection{The $\protect\xi _{\protect\alpha \protect\beta }$ functions}

We present a set of functions which characterizes three point functions in
one loop calculations. Their most general form reads \cite{ori} 
\begin{equation}
\xi _{\alpha \beta }(\mu _{1}^{2},\mu _{2}^{2},\mu
_{3}^{2},p^{2},q^{2})=\int_{0}^{1}dz\int_{0}^{1-z}dy\frac{z^{\alpha
}y^{\beta }}{Q(y,z,\mu _{1}^{2},\mu _{2}^{2},\mu _{3}^{2},p^{2},q^{2})},
\label{ksi}
\end{equation}
where $\mu _{1}^{2},\mu _{2}^{2}$ and $\mu _{3}^{2}$ stand for mass
parameters, $p^{2},q^{2}$ external momenta and $\alpha ,\beta \geq 0.$ $Q$
is defined as 
\begin{eqnarray}
Q(y,z,\mu _{1}^{2},\mu _{2}^{2},\mu _{3}^{2},p^{2},q^{2})
&=&p^{2}y(1-y)+(\mu _{1}^{2}-\mu _{2}^{2})y+q^{2}y(1-y)+ \\
&&+(\mu _{1}^{2}-\mu _{3}^{2})z-\mu _{1}^{2}-2(p\cdot q)yz.  \label{q}
\end{eqnarray}
When $\alpha =\beta =0$, (\ref{ksi}) reduces to a Spence Function \cite{pork}%
. Also, except for $\alpha =\beta =0$ the $\xi _{\alpha \beta }$ functions
can be reduced to the $Z_{\alpha }$ functions \cite{ori}. For instance, the
following identities involving $Z_{\alpha }$\ and $\xi _{\alpha \beta }$\
functions are very useful in proving Ward Identities \cite{gobbjp}: 
\begin{eqnarray}
q^{2}\xi _{10}+p\cdot q\xi _{01} &=&\frac{1}{2}[-Z_{0}(\mu _{1}^{2},\mu
_{2}^{2},p^{2})+ \\
&&-Z_{0}(\mu _{2}^{2},\mu _{2}^{2},(p-q)^{2})+ \\
&&-(q^{2}+\mu _{1}^{2}-\mu _{2}^{2})\xi _{00}]  \label{r1}
\end{eqnarray}
\begin{eqnarray}
q^{2}\xi _{20}+p\cdot q\xi _{11} &=&\frac{1}{2}[\frac{1}{2}Z_{0}(\mu
_{2}^{2},\mu _{2}^{2},(p-q)^{2})-Z_{1}(\mu _{2}^{2},\mu _{2}^{2},(p-q)^{2})+
\\
&&+\frac{3}{2}(q^{2}+\mu _{1}^{2}-\mu _{2}^{2})\xi _{10}+\frac{1}{2}%
(p^{2}+\mu _{1}^{2}-\mu _{2}^{2})\xi _{01}+ \\
&&-(\frac{1}{2}+\mu _{1}^{2}\xi _{00})]  \label{r2}
\end{eqnarray}
\begin{eqnarray}
q^{2}\xi _{11}+p\cdot q\xi _{02} &=&\frac{1}{2}[Z_{1}(\mu _{2}^{2},\mu
_{2}^{2},(p-q)^{2})-Z_{0}(\mu _{2}^{2},\mu _{2}^{2},(p-q)^{2})+ \\
&&+Z_{1}(\mu _{1}^{2},\mu _{2}^{2},p^{2})+\frac{1}{2}(q^{2}+\mu _{1}^{2}-\mu
_{2}^{2})\xi _{01}].  \label{r3}
\end{eqnarray}
The asymptotic limit of the $\xi _{\alpha \beta }$\ functions can be
calculated with the help of (\ref{lim}) and with the asymptotic limit of the
Spence Function.

As an example, consider the one loop correction for the quantum
electrodynamics vertex 
\begin{equation}
-ie\Gamma _{\nu }(p,q)=-e^{3}\int_{\Lambda }\frac{d^{4}k}{(2\pi )^{4}}\frac{%
\gamma _{\mu }(\gamma ^{\mu }p_{\mu }-\gamma ^{\mu }k_{\mu }+m)\gamma _{\nu
}(\gamma ^{\mu }q_{\mu }-\gamma ^{\mu }k_{\mu }+m)\gamma ^{\mu }}{%
[(p-k)^{2}-m^{2}][(q-k)^{2}-m^{2}](k^{2}-\mu ^{2})}
\end{equation}
where $p$ and $q$\ are external momenta and $\mu $\ an infrared cut-off.
Using the IRT one gets \cite{ri0} 
\begin{equation}
-ie\Gamma _{\nu }(p,q)=e^{3}\frac{-i}{(4\pi )^{2}}\left[ 4m(p_{\nu }+q_{\nu
})-\gamma _{\nu }(p^{2}+q^{2})-2\gamma ^{\mu }p_{\mu }\gamma _{\nu }\gamma
^{\beta }q_{\beta }\right] \xi _{00}+
\end{equation}
\begin{equation}
+e^{3}\frac{i8m}{(4\pi )^{2}}(p_{\nu }\xi _{10}+q_{\nu }\xi _{01})+
\end{equation}
\begin{equation}
-e^{3}\frac{i}{(4\pi )^{2}}\left[ \gamma _{\nu }(p^{\mu }+q^{\mu
})-2(q_{\beta }\gamma ^{\beta }\gamma ^{\mu }\gamma _{\nu }+p_{\delta
}\gamma ^{\delta }\gamma _{\nu }\gamma ^{\mu })\right] (p_{\mu }\xi
_{10}+q_{\mu }\xi _{01})+
\end{equation}
\begin{equation}
+e^{3}\frac{i}{(4\pi )^{2}}\left[ 2\gamma _{\nu }F+4\gamma ^{\mu }(p_{\mu
}p_{\nu }\xi _{20}+q_{\mu }q_{\nu }\xi _{02}+(p_{\mu }q_{\nu }+p_{\nu
}q_{\mu })\xi _{11}\right] +
\end{equation}
\begin{equation}
+e^{3}\frac{i}{(4\pi )^{2}}\left[ Z_{0}(\kappa
^{2},m^{2},p^{2})+Z_{0}(\kappa ^{2},m^{2},q^{2})\right] -\gamma _{\nu
}e^{3}I_{\ell }(m^{2}),  \label{cv}
\end{equation}
and 
\begin{equation}
\xi _{\alpha \beta }=\xi _{\alpha \beta }(\kappa ^{2},m^{2},p^{2},q^{2}),
\end{equation}
\begin{equation}
F=F(\kappa ^{2},m^{2},p^{2},q^{2})\equiv \int_{0}^{1}dz\int_{0}^{1-z}dy\ln
\left( \frac{Q(y,z,\kappa ^{2},m^{2},p^{2},q^{2})}{-m^{2}}\right)
\end{equation}
and $Q$ defined by (\ref{q}). With the aid of (\ref{r1}), (\ref{r2}) and (%
\ref{r3}) the 1-loop (\ref{cv}) finite parts could be reduced to $Z_{\alpha
} $ and $\xi _{00}$ functions. In the next section we will briefly review
some aspects of renormalization group equation and show how a mass scale
enters in the definition of renormalized parameters when IRT is used.

\section{The Renormalization Group and the Implicit Regularization Technique}

The RG techniques were originally introduced by Gell-Mann and Low \cite{4}
as a way of dealing with the large logarithms that may break down
perturbation procedures \cite{3}. Let $\Gamma (E,g,m)$ be a physical
amplitude that depends on an over-all energy scale $E$, on a dimensionless
coupling constant $g$ and a mass called $m$. If $\Gamma $ has dimensionality 
$[mass]^{D}$ then simple dimensional analysis tells us that 
\begin{equation}
\Gamma (E,g,m)=E^{D}\Gamma (1,g,\frac{m}{E}).
\end{equation}
In the limit $E\rightarrow \infty ,$ we might expect the simple power
behavior 
\begin{equation}
\Gamma (E,g,m)\rightarrow E^{D}\Gamma (1,g,0).
\end{equation}
Instead of this simple power behavior, in perturbation calculations the
factor $E^{D}$\ is found to be accompanied by powers of $\ln (E/m)$, which
can enter as $E\rightarrow \infty $ with fixed $m$ only if $\Gamma ,$ at
fixed $E,$ becomes singular as $m\rightarrow 0.$

Consider now a physical amplitude $\Gamma (E,g_{\mu },m_{\mu },\mu )$ that
depends on dimensionless coupling $g_{\mu }$ and mass $m_{\mu },$ defined 
{\em by renormalization conditions} on a sliding energy scale $\mu $ . We
define $g_{\mu }$\ in such a way that, at least for 
\begin{equation}
\mu >>m_{\mu },  \label{asymp}
\end{equation}
$g_{\mu }$ has no dependence on the scale $m_{\mu }$\ of the mass of the
theory. Again simple dimensional analysis tell us that 
\begin{equation}
\Gamma (E,g,m,\mu )=E^{D}\Gamma (1,g_{\mu },\frac{m_{\mu }}{E},\frac{\mu }{E}%
).  \label{gam}
\end{equation}
Since $\mu $\ is arbitrary, we can choose $\mu =E.$ Thus 
\begin{equation}
\Gamma (E,g,m,\mu )=E^{D}\Gamma (1,g_{E},0,1).
\end{equation}
This has no zero mass singularities because, by construction, $g_{E}$ does
not depend on $m_{\mu }$ for $E>>m_{\mu },$ so there are no large logarithms
and we can use perturbation theory to calculate $\Gamma $\ in terms of $%
g_{E} $\ as long as $g_{E}$ itself remains small\footnote{$g_{E}$ is the
finite coupling defined in a conventional renormalization point.}. Consider
eq. (\ref{gam}), an $n$ point amputated Green's function 
\begin{equation}
\Gamma (E,g,m,\mu )\equiv \Gamma _{\mu }^{(n)}(p_{1},...,p_{n},m_{\mu
},g_{\mu },\mu )
\end{equation}
obtained from a bare $n$ point amputated Green's function via multiplicative
renormalization 
\begin{equation}
\Gamma _{B}^{(n)}(p_{1},...,p_{n},m_{B},g_{B},\Lambda )=Z_{\phi _{\mu }}^{-%
\frac{n}{2}}(\frac{\Lambda }{\mu },\frac{m_{\mu }}{\mu },g_{\mu })\Gamma
_{\mu }^{(n)}(p_{1},...,p_{2},m_{\mu },g_{\mu },\mu ),  \label{aux}
\end{equation}
where $p_{1},...,p_{n}$ stand for $n$ external momenta and $\Lambda $ for an
ultraviolet cut-off. Imposing invariance of the LHS of (\ref{aux}) with
respect to $\mu $, one gets the renormalization group (RG) equation \cite{4} 
\cite{5} 
\begin{equation}
\left( \mu \frac{\partial }{\partial \mu }+\beta _{\mu }\frac{\partial }{%
\partial g_{\mu }}+\gamma _{m_{\mu }}\frac{\partial }{\partial m_{\mu }}%
-n\gamma _{\mu }\right) \Gamma _{\mu }^{(n)}=0  \label{egr}
\end{equation}
where 
\begin{equation}
\beta _{\mu }\equiv \mu \frac{\partial g_{\mu }}{\partial \mu },
\label{beta}
\end{equation}
\begin{equation}
\gamma _{m_{\mu }}\equiv \mu \frac{\partial m_{\mu }}{\partial \mu }
\label{gamma}
\end{equation}
and 
\begin{equation}
\gamma _{\mu }\equiv \frac{\mu }{2}\frac{d}{d\mu }\ln (Z_{\phi _{\mu }})
\label{gm}
\end{equation}

As stated in the introduction, in order to properly define the parameters of
a theory, one must specify renormalization conditions. These conditions
define the values of the Green's functions and, in the process, remove all
ultraviolet divergences. But the values of the renormalized parameters are
defined in {\em specific external momenta}. This simple fact is fundamental
to understand how an arbitrary mass scale enters in the definition of
renormalized parameters when handling amplitudes using the IRT. According to
this technique the (non-polynomial) external momenta dependence of the
amplitude is contained in its finite parts, duly separated in terms of the $%
Z_{\alpha }$ and/or $\xi _{\alpha \beta }$\ functions. Thus, in dealing with 
$\Gamma $ using IRT we find that the one loop renormalized parameters $%
g_{\mu },m_{\mu }$ and $\phi _{\mu }$ are given in terms of the $Z_{\alpha }$
and/or $\xi _{\alpha \beta }$ functions, which have explicit dependence on
external momenta. Hence the renormalization group coefficients (\ref{beta}),
(\ref{gamm}) and (\ref{gam}) can be directly obtained from derivatives of
the systematized $Z_{\alpha }$ and/or $\xi _{\alpha \beta }$ functions
relative to the external momenta.

Consider the $\lambda \varphi _{4}^{4}$ coupling (\ref{lamb}) evaluated in a
sliding scale $\mu $%
\begin{equation}
\lambda _{\mu }=-\lambda -\frac{3}{2}\frac{1}{(4\pi )^{2}}%
Z_{0}(m^{2},m^{2},\mu ^{2})\lambda ^{2}+O(\lambda ^{3}).  \label{lambm}
\end{equation}
The $\beta $-function (\ref{beta}) can be evaluated directly from its
definition 
\begin{equation}
\beta _{\mu }\equiv \mu \frac{\partial }{\partial \mu }\left[ -\lambda -%
\frac{3}{2}\frac{1}{(4\pi )^{2}}Z_{0}(m^{2},m^{2},\mu ^{2})\lambda
^{2}+O(\lambda ^{3}).\right] ,
\end{equation}
notice that the $\mu $-dependence of the coupling resides {\em only} on the $%
Z_{0}$ function. Hence 
\begin{equation}
\beta _{\mu }=\frac{3}{2}\frac{1}{(4\pi )^{2}}\left[ \int_{0}^{1}dz\frac{\mu
^{2}z(1-z)}{\mu ^{2}z(1-z)-m_{\mu }^{2}}\right] \lambda ^{2}+O(\lambda ^{3}).
\label{bna}
\end{equation}
Note that the result (\ref{bna}) belongs to the non-asymptotic region.
Taking the limit $\mu >>m_{\mu }$ yields 
\begin{equation}
\beta _{\mu }=\frac{3}{16\pi ^{2}}\lambda ^{2}+O(\lambda ^{3}),
\end{equation}
which is the standard one loop result to $\lambda \varphi _{4}^{4}$ theory.

The same lines of reasoning apply to quantum electrodynamics. The $\beta $%
-function calculation could be performed from its definition (\ref{beta})
with the coupling given by (\ref{cv}). This would lead to quite lenghty
calculations, fortunately avoidable by the Ward Identity 
\begin{equation}
z_{1}=z_{2}
\end{equation}
which implies 
\begin{equation}
e_{B}=\frac{e}{\sqrt{z_{3}}}
\end{equation}
or, in terms of the fine-structure constant $\alpha ,$ 
\begin{equation}
\alpha _{B}\equiv \frac{\alpha }{z_{3}}.  \label{alfa}
\end{equation}
Imposing $\mu $ invariance on eq. (\ref{alfa}) yields 
\begin{equation}
\beta =\frac{1}{z_{3}}\mu ^{2}\frac{\partial z_{3}}{\partial \mu ^{2}}\alpha
+O(\alpha ^{2}).  \label{bet}
\end{equation}
By the IRT the $z_{3}$\ finite part reads \cite{ri0} 
\begin{equation}
(z_{3})_{fin}=1-\frac{\alpha }{3\pi }\left[ \left( 1+\frac{2m^{2}}{q^{2}}%
\right) Z_{0}(m^{2},m^{2},q^{2})+\frac{1}{3}\right] ,  \label{z3}
\end{equation}
and imposing (\ref{fos}), (\ref{z3}) is evaluated in the external momentum $%
q^{2}=\mu ^{2}$. Again, that is how the sliding scale $\mu $ enters in the
IRT. In other words, in the IRT, the sliding scale $\mu $\ is directly
related to the renormalization condition. A straightforward calculation
yields the well known 1-loop asymptotic quantum electrodynamics $\beta $%
-function\footnote{%
This known result is valid in the region $\mu >>m$ .} 
\begin{equation}
\beta =\frac{2}{3\pi }\alpha ^{2}+O(\alpha ^{3}).
\end{equation}

\section{The Asymptotic Region and Connection Between Subtraction Schemes}

In perturbation theory, the dependence of Green functions on massive
parameters is expressed by two differential equations. First, the
Callan-Symanzik (CS) equation that describes the breaking of the
dilatational invariance under rescaling in the momenta \cite{a}: 
\begin{equation}
\left( m\frac{\partial }{\partial m}+\mu \frac{\partial }{\partial \mu }%
+\beta \frac{\partial }{\partial g}-n\gamma \right) \Gamma ^{(n)}(\varphi
)=\alpha \int [-m^{2}\varphi ^{2}]_{2}\Gamma ^{(n)}(\varphi )  \label{cse}
\end{equation}
where $m$ and $\mu $ are mass parameters, $g$ the coupling and $\gamma $ the
anomalous dimension. The other equation is the RG equation derived above (%
\ref{egr}): 
\begin{equation}
\left( \mu \frac{\partial }{\partial \mu }+\beta \frac{\partial }{\partial g}%
+\gamma _{m}\frac{\partial }{\partial m}-n\gamma \right) \Gamma
^{(n)}(\varphi )=0.  \label{esp}
\end{equation}
As we showed, (\ref{esp}) describes the invariance of Green functions under
the renormalization group transformations. The original ideas that lead to (%
\ref{esp}) stem from Stueckelberg and Petermann \cite{5} and Gell-Mann and
Low \cite{4}. Mass independent $\beta $-functions in massive theories
indicate the fact that the renormalization group transformations are
restricted to the asymptotic region \cite{kraus}. The condition for mass
independence in the asymptotic region is the existence of a CS equation of
the same form of the RG equation, i.e. the known massless differential
operators of the CS equation must be the same operators of the RG equation,
evaluated in the large momenta region. It was proved for the massive $%
\lambda \varphi ^{4}$ theory (though the proof is quite general \cite{kraus}%
) that the Minimal Subtraction $(MS)$, Modified Minimal Subtraction $(%
\overline{MS)}$ \cite{ms}\ and the BPHZL \cite{bphz}\ schemes have
normalization properties in the asymptotic region: 
\begin{equation}
\lim_{p^{2}>>m^{2}}\frac{\partial }{\partial p^{2}}\Gamma ^{(2)}(p^{2}=\mu
^{2})=\sum_{n=0}^{\infty }a_{(2)n}^{n}\lambda ^{n}  \label{derlim}
\end{equation}
\begin{equation}
\lim_{p^{2}>>m^{2}}\Gamma ^{(4)}(p_{i}^{2}=\mu ^{2},p_{i}.p_{j}=-\frac{\mu
^{2}}{3})=\sum_{n=0}^{\infty }a_{(4)n}^{n}\lambda ^{n+1}  \label{coulim}
\end{equation}
where the $a_{(k)n}^{n}$ are mass independent coefficients. Hence, in all
those three schemes the $\beta $-functions and the $\gamma $-functions of
the CS equation and RG equation are the same and mass independent. As we saw
above, the result (\ref{bna}) is in the non-asymptotic region and in the
limit $\mu >>m_{\mu }$ (\ref{lambm}) yields using (\ref{lim}) 
\begin{equation}
\lambda _{\mu }=-\lambda +\left[ \frac{1}{(4\pi )^{2}}\frac{3}{2}\ln (\frac{%
\mu ^{2}}{m^{2}})\right] \lambda ^{2}+O(\lambda ^{3}).  \label{lim2}
\end{equation}
It is interesting to compare (\ref{lim2}) to general 1-loop results of $MS$, 
$\overline{MS}$ and the $BPHZL$\ schemes \cite{kraus}: 
\begin{equation}
MS:\lambda _{\mu }=-\lambda +\left[ Z_{0}(\kappa ^{2},m^{2})+\frac{1}{(4\pi
)^{2}}\frac{3}{2}\ln (\frac{\mu ^{2}}{m^{2}})+\ln 4\pi -\gamma _{E}\right]
\lambda ^{2}+O(\lambda ^{3})  \label{ms}
\end{equation}
\begin{equation}
\overline{MS}:\lambda _{\mu }=-\lambda +\left[ Z_{0}(\kappa ^{2},m^{2})+%
\frac{1}{(4\pi )^{2}}\frac{3}{2}\ln (\frac{\overline{\mu }^{2}}{m^{2}})%
\right] \lambda ^{2}+O(\lambda ^{3})  \label{msb}
\end{equation}
\begin{equation}
BPHZL:\lambda _{\mu }=-\lambda +\left[ Z_{0}(\kappa ^{2},m^{2})+\frac{1}{%
(4\pi )^{2}}\frac{3}{2}\ln (-\frac{4\mu ^{2}}{3m^{2}})\right] \lambda
^{2}+O(\lambda ^{3}).  \label{bph}
\end{equation}
Where we denote the renormalization point according to the general
convention $\mu $ and $\overline{\mu }$ respectively. It is important to
observe that the renormalization conditions that yield the expansions above
in the finite conventional coupling $\lambda $ are different from the one we
adopted. In (\ref{ms}), (\ref{msb}) and (\ref{bph}) the parameter $\lambda $
is evaluated in the Euclidean symmetric point $(p^{2}<0)$: 
\begin{equation}
p_{i}^{2}=\kappa ^{2}
\end{equation}
and 
\begin{equation}
p_{i}p_{j}=-\frac{\kappa ^{2}}{3}.
\end{equation}
As stated above, the choice of this point has not physical relevant
consequences. Anyway, contact with results the (\ref{ms}), (\ref{msb}) and (%
\ref{bph}) can be done with the help of the identity (\ref{re}). Taking $%
\kappa ^{2}=0$ in (\ref{ms}), (\ref{msb}) and (\ref{bph}) yields 
\begin{equation}
MS:\lambda _{\mu }=-\lambda +\left[ \frac{1}{(4\pi )^{2}}\frac{3}{2}\ln (%
\frac{\mu ^{2}}{m^{2}})+\ln 4\pi -\gamma _{E}\right] \lambda ^{2}+O(\lambda
^{3}),
\end{equation}
\begin{equation}
\overline{MS}:\lambda _{\mu }=-\lambda +\left[ \frac{1}{(4\pi )^{2}}\frac{3}{%
2}\ln (\frac{\overline{\mu }^{2}}{m^{2}})\right] \lambda ^{2}+O(\lambda
^{3}),
\end{equation}
\begin{equation}
BPHZL:\lambda _{\mu }=-\lambda +\left[ \frac{1}{(4\pi )^{2}}\frac{3}{2}\ln (-%
\frac{4\mu ^{2}}{3m^{2}})\right] \lambda ^{2}+O(\lambda ^{3}),
\end{equation}
and comparing those results with (\ref{lim2}) we see indeed that the IRT
applied to the $\lambda \varphi ^{4}$ yields the same asymptotic expressions
as the known schemes above. Again, we would like to stress that the
connection between sliding scales and renormalized parameters is quite
natural in the IRT, since it is realized by imposing renormalization
conditions.

\section{Relationship between dimensional, differential and implicit
renormalizations}

In this section we will show how DR, differential regularization (DFR) \cite
{dune} and IRT are related regarding the appearance of a renormalization
scale. Such comparision is interesting since DR is widely used for analysing
renormalizable QFT (particularly those involving gauge symmetry) whereas DFR
is an elegant framework which, as well as IRT, does not recourse to
analytical continuation on the space time dimension. The idea behind DFR is
to redefine products of Green's functions in the Euclidean (position) space
as proper distributions: the singularities at coincident points (which have
no Fourier transform) are expressed as derivatives of less singular terms
(which do have Fourier transforms). We follow \cite{dune}.

Consider the identity: 
\begin{equation}
\left| x\right| ^{-p}=\frac{\Box \left| x\right| ^{-p+2}}{(-p+2)(d-p)}.
\label{dal1}
\end{equation}
For $p=d$ \ we can not use (\ref{dal1}) because of the pole. According to
the DFR rules, we must instead substitute 
\begin{equation}
\left| x\right| ^{-p}\mid _{reg}:=\frac{1}{2(2-d)}\Box \frac{\ln M^{2}\left|
x\right| ^{2}}{\left| x\right| ^{d-2}},  \label{dal2}
\end{equation}
which holds when $\left| x\right| \neq 0$ and the dependence on an arbitrary
mass scale $M$ appears for dimensional reasons. It plays the role of scale
in the Callan-Symanzik renormalization group equation.

To make contact with RD we can use identity (\ref{dal1}) by extending $d$ to 
$d-r\epsilon $ where it is well defined to write: 
\begin{eqnarray}
\mu ^{r\epsilon }\left| x\right| ^{-d+r\epsilon } &=&\frac{1}{\epsilon }\mu
^{r\epsilon }\frac{1}{r(2-d+r\epsilon )}\Box \left| x\right| ^{d-2}\ln
M^{2}\left| x\right| ^{-d+r\epsilon +2}  \nonumber \\
&=&\frac{1}{\epsilon }\frac{4\pi ^{d/2}}{r(2-d+r\epsilon )\Gamma (d/2-1)}%
\delta ^{(d)}(x)+\frac{1}{2(2-d)}\Box \frac{\ln \mu ^{2}\left| x\right| ^{2}%
}{\left| x\right| ^{d-2}}+O(\epsilon ).  \label{dal3}
\end{eqnarray}

Now we can clearly see that finite (no counterterms) part of (\ref{dal3}) is
identical to DFR after subtracting the infinite and a finite $O(\epsilon
^{0})$ counterterms represented by the delta function and identifying $\mu $
with $M.$

As a matter of illustration consider the one-loop four point function of $%
\varphi ^{4}$ theory. In DR it reads

\begin{equation}
\Gamma (p^{2},m^{2})=\frac{i\lambda ^{2}\mu ^{\epsilon }}{16\pi ^{2}\epsilon 
}-\frac{i\lambda ^{2}\mu ^{\epsilon }}{32\pi ^{2}}\left\{ \gamma
+\int_{0}^{1}dz\ln [\frac{p^{2}z(1-z)-m^{2}}{4\pi \mu ^{2}}]\right\} .
\label{dal4}
\end{equation}
By defining counterterms to subtract the pole and the term proportional to $%
\gamma $ enables us to write 
\begin{equation}
\Gamma _{RD}^{R}(p^{2},m^{2})=\frac{i\lambda ^{2}}{32\pi ^{2}}\left\{ \ln (%
\frac{m^{2}}{4\pi \mu ^{2}})-Z_{0}(p^{2},m^{2})\right\} .  \label{dal5}
\end{equation}

According to the DFR rules $\Gamma ^{R}(p^{2},m^{2})$ is written as \cite
{torre} 
\begin{equation}
\Gamma _{DFR}^{R}(p^{2},m^{2})=\frac{i\lambda ^{2}}{32\pi ^{2}}\left\{ \ln (%
\frac{m^{2}}{M^{2}})-Z_{0}(p^{2},m^{2})\right\} ,  \label{dal6}
\end{equation}
from which is clear the equivalence of (\ref{dal5}) and (\ref{dal6})
identifying $M^{2}=4\pi \mu ^{2}.$ Using now the IRT rules we can write 
\begin{equation}
\Gamma _{IRT}^{R}(p^{2},m^{2})=\frac{\lambda ^{2}}{2}\left\{ I_{\ell
}(m^{2})-\frac{i}{4\pi ^{2}}Z_{0}(p^{2},m^{2})\right\} .  \label{dal7}
\end{equation}
Taking 
\begin{equation}
I_{\ell }(m^{2})=I_{\ell }(\eta ^{2})+\frac{i}{4\pi ^{2}}\ln \frac{\eta ^{2}%
}{m^{2}}
\end{equation}
where the second term on the RHS of the equation above parametrizes a finite
arbitrary counterterms, into (\ref{dal7}) and defining a counterterms to
subtract $I_{\ell }(\eta ^{2})$ leads to 
\begin{equation}
\Gamma _{IRT}^{R}(p^{2},m^{2})=\frac{i\lambda ^{2}}{32\pi ^{2}}\left\{ \ln (%
\frac{m^{2}}{\eta ^{2}})-Z_{0}(p^{2},m^{2})\right\} ,
\end{equation}
making clear the connection between these three schemes.

\section{Conclusions}

In treating quantum field theory amplitudes perturbatively, a
renormalization procedure must be imposed to define, order by order, the
parameters of the theory. Such perturbative procedures are plagued by
divergences. To remove the divergences and, in the process redefine the
parameters of the theory, a subtraction scheme associated with a
regularization method must be employed. This procedure can not be performed
in a unique way since divergences are present. Thus any subtraction
algorithm must carry a parameter to accomplish the arbitrariness of this
infinite renormalization. A non trivial issue is what is the role played by
this parameter in the theory. The most general prescription is to state
renormalization conditions which define the values of the Green's functions
in a sliding scale and, in the process, remove all ultraviolet divergences.
When the dimensional regularization is used, an arbitrary mass parameter
must be introduced in order to keep the coupling dimensionality. This
parameter is linked non-trivially to the parameter introduced via
renormalization conditions and thus it is used for renormalization group
purposes. Employing the BPHZL scheme, no use of an explicitly cut-off is
required {\em a priori}. But the for practical purposes, very often the
dimensional regularization is implemented in intermediary steps and hence
breaking the 4-space dimensionality. Moreover, in IRT we explicitly
construct the counterterms without changing the structure of the integrand.
Using the IRT the role of renormalization conditions concerning the
introduction of a sliding scale is straightforward: the arbitrariness of the
sliding scale is directly related to the unavoidable arbitrariness in the
process of separating finite and divergent contributions to physical
amplitudes. Moreover, as we have shown, the finite 1-loop parts are
systematized in a small set of \ functions, namely the $Z_{\alpha }$ and $%
\xi _{\alpha \beta }$ \ functions. The renormalization conditions act
directly in the $Z_{\alpha }$ and $\xi _{\alpha \beta }$ \ functions which
contain all the arbitrariness of the process. The connection of the
systematized\ functions with renormalization group results show that IRT
results are calculated in non-asymptotic region and agree with standard
calculations in the large momenta region. The extension of our approach to
non-abelian gauge theories where BPHZ fails to preserve the relevant
Slavnov-Taylor identities (also in the finite part) is presently under
study. Since we do not change the structure of the integrand and keep the
arbitrariness expressed by differences between divergent integrals to be
fixed on physical grounds related to momentum routing invariance we expect
that we can fully preserve gauge invariance.

\section*{Acknowledgments}

This work was partially supported by FAPEMIG, CNPq and FAPESP. M.S. acknowledges a
fellowship from FCT - Portugal under the grant number BPD/22016-99.

\end{document}